\documentclass[prb,twocolumn,superscriptaddress]{revtex4}

\usepackage{graphicx}

\newcommand{\bra}[1]{\ensuremath{\left\langle #1\right|}}
\newcommand{\ket}[1]{\ensuremath{\left|#1\right\rangle}}

\usepackage[normalem]{ulem} 
\usepackage{color} 

\begin{document}

\title{Multi-photon spectroscopy of a hybrid quantum system}

\author{P.~Bushev}
\affiliation{Physikalisches Institut, Karlsruhe Institute of Technology, D-76128 Karlsruhe, Germany}

\author{C.~M\"uller}
\affiliation{Institut f\"{u}r Theorie der Kondensierten Materie, Karlsruhe Institute of Technology, D-76128 Karlsruhe, Germany}
\affiliation{DFG-Center for Functional Nanostructures (CFN), D-76128 Karlsruhe, Germany}

\author{J.~Lisenfeld}
\affiliation{Physikalisches Institut, Karlsruhe Institute of Technology, D-76128 Karlsruhe, Germany}

\author{J.~H.~Cole}
\affiliation{Institut f\"ur Theoretische Festk\"{o}rperphysik, Karlsruhe Institute of Technology, D-76128 Karlsruhe, Germany}
\affiliation{DFG-Center for Functional Nanostructures (CFN), D-76128 Karlsruhe, Germany}

\author{A.~Lukashenko}
\affiliation{Physikalisches Institut, Karlsruhe Institute of Technology, D-76128 Karlsruhe, Germany}

\author{A.~Shnirman}
\affiliation{Institut f\"{u}r Theorie der Kondensierten Materie, Karlsruhe Institute of Technology, D-76128 Karlsruhe, Germany}
\affiliation{DFG-Center for Functional Nanostructures (CFN), D-76128 Karlsruhe, Germany}

\author{A.~V.~Ustinov}
\affiliation{Physikalisches Institut, Karlsruhe Institute of Technology, D-76128 Karlsruhe, Germany}
\affiliation{DFG-Center for Functional Nanostructures (CFN), D-76128 Karlsruhe, Germany}

\begin{abstract}
	We report on experimental multi-photon spectroscopy of a hybrid quantum system consisting of a superconducting phase qubit coherently coupled to an intrinsic two-level defect.
	We directly probe hybridized states of the combined qubit-defect system in the strongly interacting regime,
	where both the qubit-defect coupling and the driving cannot be considered as weak perturbations.
	This regime is described by a theoretical model which incorporates anharmonic corrections, multi-photon processes and decoherence.
	We present a detailed comparison between experiment and theory and find excellent agreement over a wide range of parameters.
\end{abstract}

\maketitle

\section*{I. Introduction}

One of the fundamental tenets of quantum mechanics is that the rules that govern a system's behavior do not depend on its physical size.
Rather, it is the number of degrees of freedom a system possesses and how well the system is isolated from its environment,
which dictates how it behaves and whether it is considered a quantum or classical system.
This is most clearly seen in the plethora of different systems in which quantum mechanical phase coherence can be seen and controlled.
These include atomic (sub nm) systems such as trapped ions~\cite{BlattWineland} or optical color-centers~\cite{Wrachtrup08},
mesoscopic systems (nm-$\mu$m) such as superconducting devices~\cite{Clarke88}
or macroscopic objects (mm) such as Bose-Einstein condensates~\cite{BEC1}.

In this paper, we report on experiments with a hybrid quantum system~\cite{Zoller} consisting of two parts of different length-scale and physical nature,
namely a superconducting phase qubit ($\sim\mu$m) and an intrinsic two-level defect ($\sim$nm).
The systems are coupled via an effective dipole-dipole interaction in a regime where both the interaction between its components and the driving field are strong.
We observe effects which are not present in the usual simplified models~\cite{KofmanPRB06}, including multi-photon transitions and weak violation of selection rules due to higher order corrections to the anharmonic potential.


When studying superconducting qubits spectroscopically, one often finds clear signatures of level anticrossings
at certain frequencies~\cite{Simmonds04, Cooper04, Martinis:05, Lupascu09, Jurgen09, Palomaki:10}.
These anticrossings indicate intrinsic, microscopic two-level systems (TLS) being coupled coherently to the qubit circuit.
In general, ensembles of two-level microscopic defects are believed to be responsible for loss in a wide variety of systems,
including phase- and flux-based superconducting qubits and even nanomechanical oscillators~\cite{Arcizet:2009, Venkatesan:2009},
as well as more general effects in amorphous and spin-glass systems\cite{Esquinazi:1998}.
In superconducting systems, the exact microscopic nature of these defect two-level systems remains unknown.
It has been shown that they are coherent two-level systems, and due to their relatively long coherence times
can potentially be used to store and retrieve quantum information~\cite{Neeley08}.
In general, two-level defects are considered detrimental to the qubits operation,
since they introduce additional channels of decoherence~\cite{Schriefl06,Clemens09} and therefore, acquiring a more in-depth understanding of their nature is essential.
It has been suggested that these TLSs are formed by charged microscopic defects, located inside the insulating tunnel barrier of the Josephson junction~\cite{Martinis:05, Martin:05} and interacting with its electric field.
In this case, the coupling to the qubit circuit can be written as $\propto \vec{d}_{f} \cdot \vec{E}$,
where $\vec{d}_{f}$ is the dipole moment of the TLS and $\vec{E}$ is the electric field across the Josephson junction (JJ).
Another suggested mechanism for qubit-defect coupling is that the state of the TLS modulates the transparency of the junction~\cite{Simmonds04, Martin:05}
and therefore its critical current $I_{C}$. In this case two-level defects could be formed by Andreev bound states~\cite{Sousa:2009} or Kondo-impurities~\cite{Faoro:2007, Faoro:2008}.
Since the form of the potential in the qubit circuit depends on the strength of the critical current, a modulation of $I_{C}$ will always result in a direct change of the level splitting in the circuit.
Since we do not observe signatures of longitudinal coupling in our experiments (as has also been shown in flux qubits~\cite{Lupascu09}),
we rule out this coupling mechanism and take the TLS to be a charged defect.

\section*{II. Experiment}

\begin{figure}[htb]
    	\includegraphics[width=.7\columnwidth]{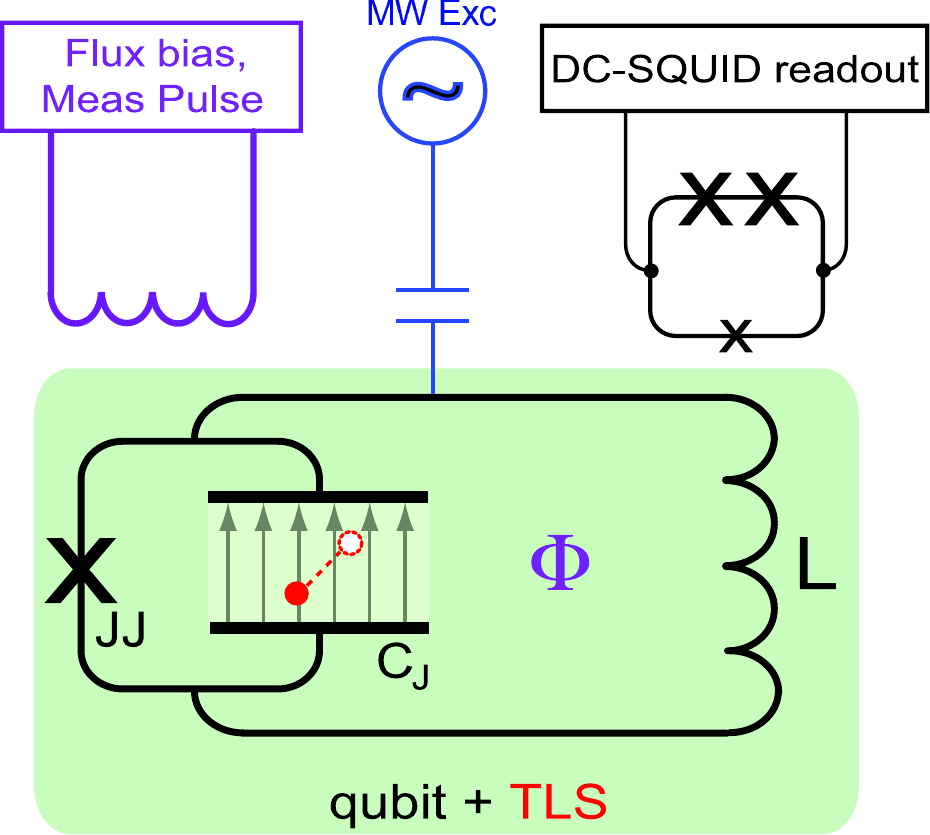}
    	\caption{Sketch of the experimental setup and circuit of the phase qubit with the TLS residing inside the qubits Josephson junction.}
	\label{fig:Circuit}
\end{figure}

We measured a superconducting phase qubit, consisting of a single Josephson junction embedded in a superconducting loop (Fig.~\ref{fig:Circuit}).
The experiments reported here are performed with the qubit circuit biased such that the frequency of its lowest lying transition is close to the resonance frequency of a particular TLS.
The studied TLS defect has a level splitting of $\epsilon_{f}/h = 7.845$ GHz and is coupled to the qubit with a coupling strength $v_{\perp}/h=21$ MHz.

The JJ is shunted with a capacitor $C_{S}$ of $800$ fF, which is needed to tune the circuit to the desired frequency range. Its critical current is $I_c = 1.1$~$\mu$A and the inductance of the loop is $L= 720$~pH. During the experiment, the chip is maintained at a temperature of approximately $40$~mK.

By applying an external magnetic flux through the loop, the shape of the potential of the quantum circuit can be tuned.
In the parameter regime of interest, the potential consists of two wells, one of which is very shallow.
Operation of the circuit takes place in this anharmonic shallow well, containing five to ten energy levels~\cite{KofmanPRB06}.
The two lowest energy states in this well form a macroscopic two-level system (our qubit).
The transition frequencies are in the microwave range and the state of the system can be manipulated via capacitively coupled microwaves.
In order to readout the state of the qubit, the potential barrier between the two wells is lowered to allow the excited states of the shallow well to tunnel into the deeper well.
States in the two wells differ by a large amount of persistent current through the loop,
which can be detected by measuring the switching current of a measurement SQUID located on the same chip.
The resulting escape probability is normalized to the measurement contrast defined by the qubit-SQUID coupling.
A measurement thus distinguishes between the ground state and all other, excited states of the qubit circuit and yields the overall excitation probability for the qubit
or the probability not to be in the ground state.

In the following we present data on microwave spectroscopy of the circuit. The system is driven for a sufficiently long time, $\sim 2$~$\mu$s,
so that all transient phenomena have decayed.  The state of the qubit after this pulse is measured as a function of microwave frequency and flux bias to map out the system transitions.

\section*{III. Theory}

When considering a phase qubit which is weakly driven and/or weakly coupled to a TLS,
it is common to model the qubit circuit as a two-level system and to use the rotating-wave-approximation.
For our experiment, neither of the above approximations are valid and one must use a formalism
which takes into account an arbitrary number of qubit states as well as multi-photon processes.
This allows us to describe the experiment over a wide parameter range without knowing which effects are important \emph{a priori}.
Several techniques exist for treating a driving field when the perturbation is no longer ``weak'' \cite{Wilson:07,Hausinger10,Wilson:10}.
We employ a numerical expansion in the Floquet basis~\cite{Bain:2001,Shirley:65} as this allows us to include arbitrary multi-photon processes together with decoherence.
The inclusion of higher lying states in the qubit can be achieved via direct diagonalization of the exact qubit potential,
or by using an anharmonic approximation~\cite{KofmanPRB06,Neeley08}.

The qubit is well described by a quantum anharmonic oscillator model, with the anharmonicity determined by the Josephson energy $E_{J}$~\cite{KofmanPRB06}.
In the parameter regime explored in the experiment, it is sufficient to describe the qubit by using only the lowest three energy levels.
The TLS is modeled as a two-level system with its energy splitting $\epsilon_{f}$ obtained from the spectroscopic data.
We write the coupling term between qubit and TLS as  $v_{\perp} \: \hat{p} \cdot \hat{\tau}_{x}$, where the operator $\hat{p}$ is proportional to the charge
and therefore the electric field across the qubit circuit's Josephson junction.
The operator $\hat{\tau}_x$ is the Pauli matrix in $x$-direction acting on the two-level defect, inducing transitions of the state of the TLS.
The Hamiltonian is given by
\begin{equation}
    \hat{H}_{0} = \epsilon_{01} \ket{1}\bra{1} + (\epsilon_{01} + \epsilon_{12}) \ket{2}\bra{2}
        + \frac{1}{2} \epsilon_{f} \hat{\tau}_{z} + \frac{1}{2} v_{\perp}\: \hat{p} \cdot \hat{\tau}_{x} \,.
        \label{eq:H0}
\end{equation}
\noindent
The operator $\hat{p}$ can be written in the harmonic oscillator basis as $\propto i (\hat{a}^{\dagger} - \hat{a})$,
where $\hat{a}$, $\hat{a}^\dagger$ are the usual annihilation and creation operators. In the exact qubit eigenbasis $\ket{n}$, used to write down Eq. (\ref{eq:H0}),
this operator will acquire corrections of the form $\ket{n}\bra{n+2}$.
To obtain the energy splittings $\epsilon_{01}$ and $\epsilon_{12}$ in this basis,
we solve numerically using the exact potential of the phase qubit circuit.
The state of the TLS is described by its ground state $\ket{g}$ and excited state $\ket{e}$.
The energies are defined relative to the energy of the ground state of the coupled qubit-TLS system $\ket{0g}$.
The system is coupled to an external microwave drive at frequency $\omega_{d}$ with the coupling Hamiltonian
\begin{equation}
    \hat{H}_{I} = \frac{1}{2} V_{q} \hat{p} \cos{\omega_{d} t} \,,
    \label{eq:HDriving}
\end{equation}
\noindent where the coupling constant $V_{q}$ is proportional to the microwave amplitude and we define the generalised $n$-photon Rabi frequency
$\Omega_q^{(n)} \propto V^n_q$ on resonance.

Previously, we have shown that close to resonance,
the TLS is driven by a quasi-direct driving term mediated by a second order process involving the qubit circuit's third level \cite{Jurgen09}.
As the model here already includes the third level by construction, this effect is incorporated automatically.

To describe the effects of decoherence in the system, we use  a simple Lindblad model \cite{Lindblad76,GardinerBook},
with relaxation and dephasing rates taken from independent measurements of both qubit and defect.
In numerical modeling of the system, we use the values of $T_1^{(q)} = 120$ ns and $T_1^{(f)} = 715$ ns for the relaxation rates of the qubit and defect, respectively.
Dephasing rates , determined from Ramsey fringes, are found to be $T_2^{(q)} = 90$ ns for the qubit and $T_2^{(f)} = 110$ ns for the TLS.
The dynamical equations for the reduced density matrix of the system (consisting of qubit and TLS) are given by
\begin{equation}
	\dot{\rho} = i \left[ \rho, \hat{H}_{0} \right] + \sum_{j} \Gamma_{j} \left( L_{j} \rho L_{j}^{\dagger} - \frac{1}{2} \left\{ L_{j} L_{j}^{\dagger}, \rho \right\} \right)
\end{equation}
where the sum is over all possible channels of decoherence with the respective rates $\Gamma_{j}$.
The $L_{j}$ are the operators corresponding to each decoherence channel, e.g., pure dephasing of the qubit is described by the operator $\hat{a}^{\dagger} \hat{a}$
while relaxation of the qubit is described by the anihilation operator $\hat{a}$.
This formalism in principle enables us to include an arbitrary number of qubit levels without changing the structure of the theory.

\section*{IV. Results and Discussion}

\subsection{Two-photon spectroscopy of 4-level hybrid quantum system}

\begin{center}
	\begin{figure}[htb]
	    	\includegraphics[width=.9\columnwidth]{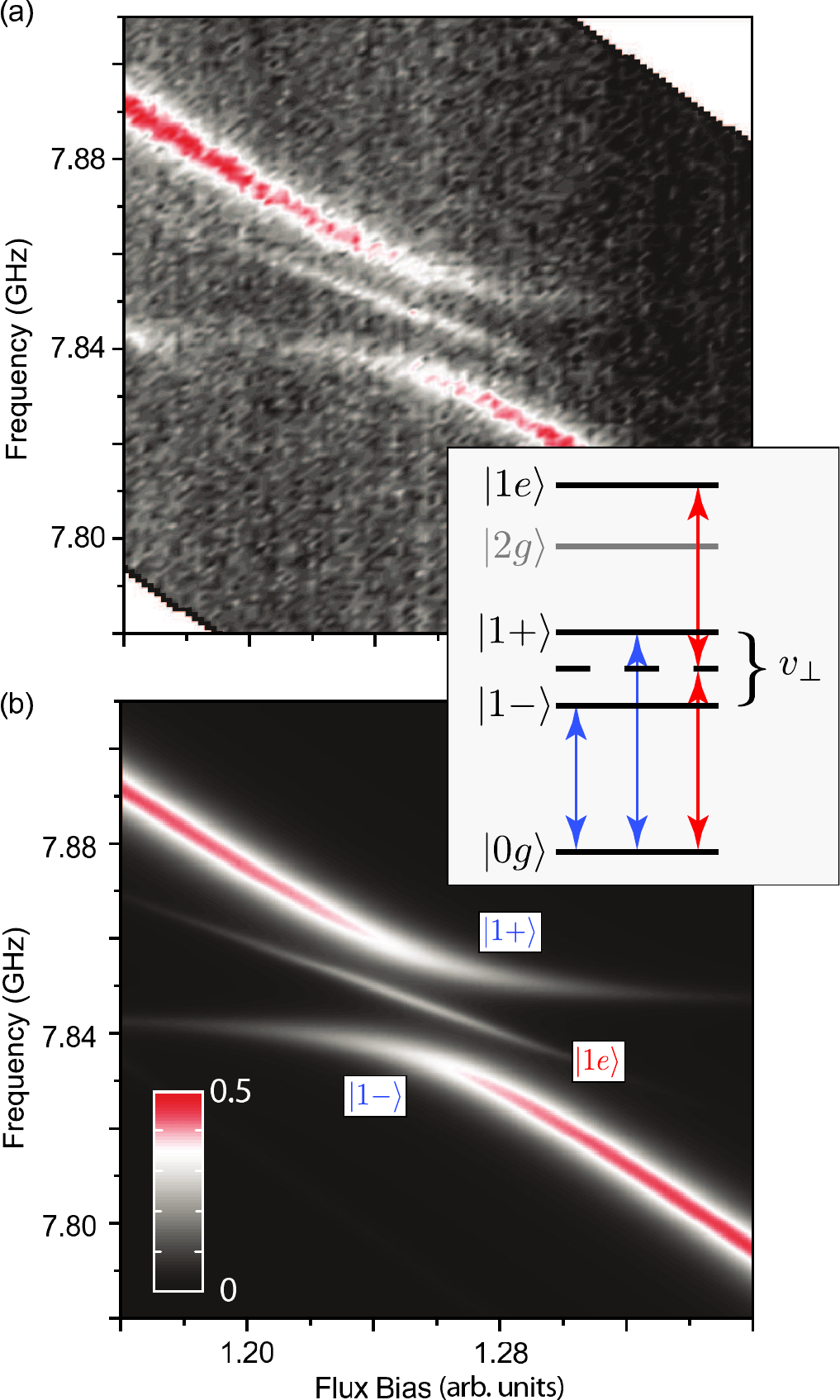}
	    	\caption{(Color online) Experimental \textbf{(a)} and theoretical \textbf{(b)} spectroscopic scans as a function of flux bias and drive frequency,
			showing both the anticrossing due to single-photon transitions and the resonance line associated with the weaker two-photon transition.
			The colorscale indicates the probability to find the qubit not in the ground state. We use the same colorscale for both plots.
			\textbf{Inset:} Energy level structure of the qubit/defect system including the hybridised states $\ket{1\pm}$, which are split by the coupling $v_{\perp}$.
			As well as the usual single-photon transitions (between $\ket{0g}$ and $\ket{1\pm}$),
			a two-photon transition is allowed between the states $\ket{1e}$ and $\ket{0g}$. 
		}
		\label{fig:Low}
	\end{figure}
\end{center}

We initially present results from driving the qubit with a relatively low microwave power (Fig.~\ref{fig:Low}),
corresponding to a qubit Rabi-frequency of $\Omega^{(1)}_q/h \approx4$~MHz. In this regime,
it is sufficient to approximate the qubit circuit by a two-level system and we can write the Hamiltonian
\begin{equation}
	\hat{H}_{0} \approx \frac{1}{2} \epsilon_{01} \hat{\sigma}_{z} + \frac{1}{2} \epsilon_{f} \hat{\tau}_{z} + \frac{1}{2} v_{\perp} \hat{\sigma}_{y} \hat{\tau}_{x}
\end{equation}
with Pauli matrices for the qubit $\sigma$ and for the defect $\tau$.
A sketch of the level structure including the relevant transitions is shown in the inset of Fig.~\ref{fig:Low}.
Here we define the hybridized eigenstates $\ket{1-} = \cos\theta \ket{0e} - \sin \theta \ket{1g}$ and $\ket{1+} = \sin\theta \ket{0e} + \cos \theta \ket{1g}$,
where $\tan 2\theta = v_{\perp} / (\epsilon_{01} - \epsilon_f)$ varies with flux bias.
The coupling term $\hat{\sigma}_{y} \hat{\tau}_{x}$ is formally equivalent to the previously described dipole-dipole coupling $\sim \vec{E} \cdot \vec{d}_{f}$.
The experimental data in Fig.~\ref{fig:Low}(a) shows a characteristic level anticrossing associated with the qubit-TLS resonance, $\epsilon_{01} \approx \epsilon_{f}$,
with coupling strength $v_{\perp}/h = 21$~MHz.

An additional spectroscopic line can be seen in the middle of the qubit-defect anticrossing.
We identify this line as a two-photon transition from the ground-state $\ket{0g}$ to the excited state $\ket{1e}$, shown by red arrows in the inset of Fig.~\ref{fig:Low}.
Similar two-photon absorption mediated by dipole-dipole coupling is observed in experiments with pairs of dye molecules at high excitation power~\cite{Vahid02}.

In Fig.~\ref{fig:Low}(b) we show the result of a theoretical treatment of the system, calculating the time dependence of the system's density matrix including decoherence.
System parameters are taken from the measurements as described above. As can be seen, the numerics reproduce the experimental data with high accuracy.

\begin{center}
	\begin{figure}[htb]
		\includegraphics[width=.8\columnwidth]{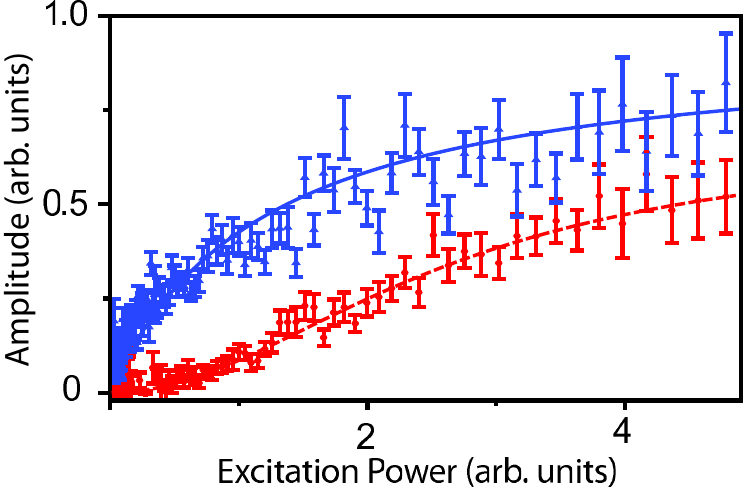}
		\caption{(Color online) Detailed power dependence of both the single (top, solid blue) and two-photon (lower, dashed red) transitions close to resonance with the defect
			(as shown in Fig.~\ref{fig:Low}).
			The amplitudes and associated uncertainties are obtained by fitting the experimental data with Lorentzian functions.
			The scaling of the saturation curves is as expected for both single- and two-photon transitions, as indicated by the fitted lines (see text).}
		\label{fig:Power}
	\end{figure}
\end{center}

The power dependence of the two-photon feature is shown in Fig.~\ref{fig:Power}, to confirm that the observed middle line results from a two-photon transition.
In this figure the amplitudes of the two-photon ($\ket{0g}\leftrightarrow \ket{1e}$) and the lower single-photon ($\ket{0g} \leftrightarrow \ket{1-}$) transitions
are plotted as a function of driving power. The data is fitted with the conventional saturation curve~\cite{Cohen:1992}
\begin{equation}
	s \propto \frac{(\Omega^{(n)}/\Gamma)^2}{[1+(\Omega^{(n)}/\Gamma)^2]},
\end{equation}
where $\Omega^{(n)}/\Gamma$ is proportional to the ratio of ($n$-photon) Rabi frequency to loss rate and $n=1,2$ is the order of the photon transition, confirming the single and two-photon nature of these two transitions, respectively.

The position of the two-photon line in the middle of the anticrossing gives an indication of the strength of a possible longitudinal coupling
$\sim \hat{\sigma}_{z} \hat{\tau}_{z}$ between qubit and TLS.
Through the presence of the two-photon line exactly in the middle of the anticrossing, we can conclude
(i) the TLS is a two-level or at least strongly anharmonic system and (ii) longitudinal coupling between qubit circuit and TLS is negligible.
Very similar features have been observed earlier in a related system, namely a superconducting flux qubit~\cite{Lupascu09},
which suggests that these strongly coupled TLS have the same origin in both flux and phase qubits, even though the degrees of freedom being manipulated are different. Additional evidence against the TLS being an harmonic object (as would result from an ensemble of two-level systems forming an effective TLS\cite{Clemens09}, for example)is provided by experiments trying to pump two excitations resonantly into the TLS\cite{HofheinzPrivate09}, similar to the protocol used in Ref.~\onlinecite{Neeley08}.

\begin{center}
	\begin{figure}[htb]
	    	\includegraphics[width=.9\columnwidth]{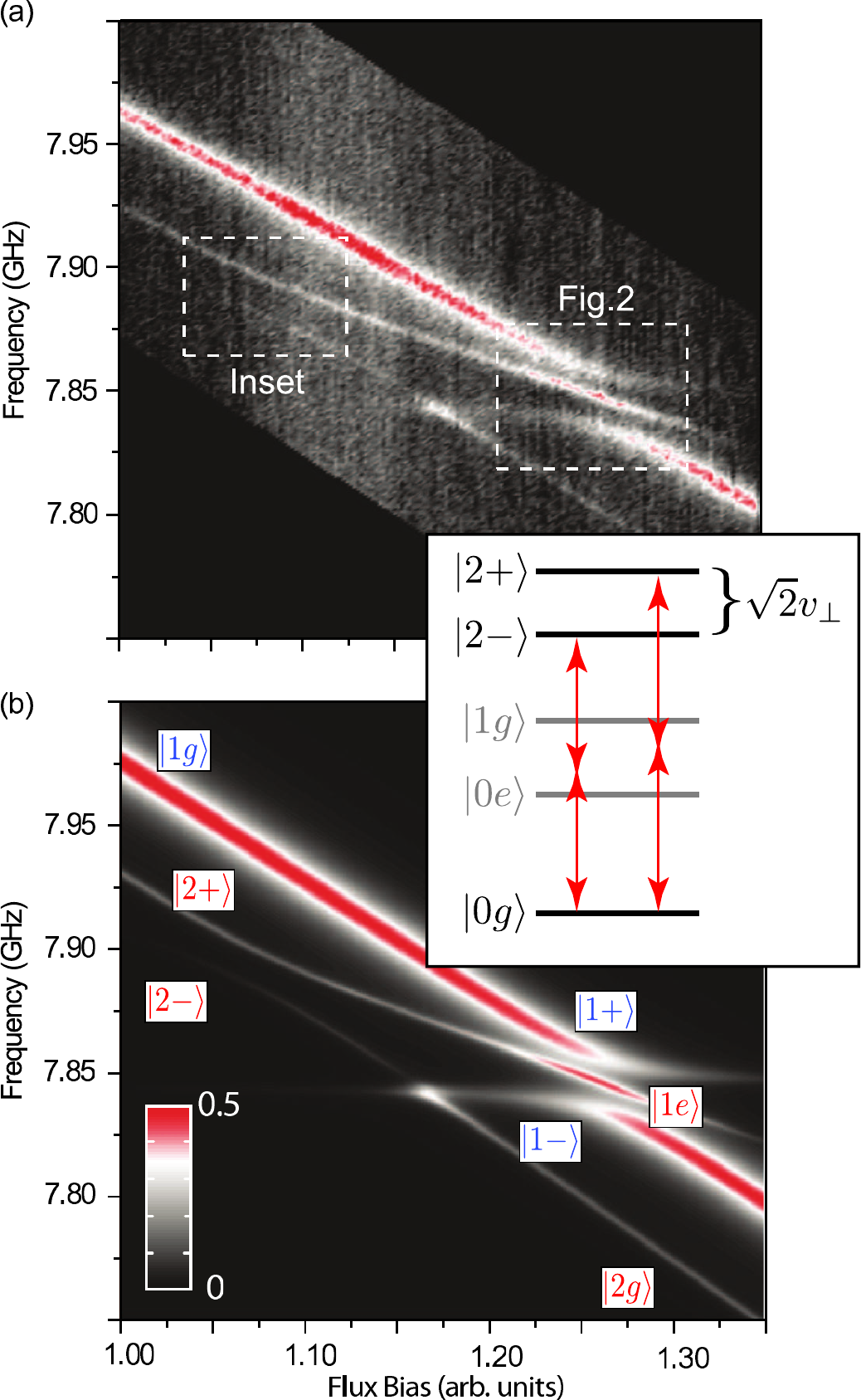}
	    	\caption{(Color online) At higher microwave powers and over a wider region of parameter space, processes involving higher lying states of the qubit become important.
			Experimental \textbf{(a)} and theoretical \textbf{(b)} spectroscopic scan as a function of flux bias,
			showing both the transitions to the hybridised states $\ket{1\pm}$ (as per Fig.~\ref{fig:Low}) and the states $\ket{2\pm}$. The colorscale is the same for both graphs.
			\textbf{Inset:} Level structure including higher lying states of the qubit including further hybridised states which can be excited via a two-photon process, as indicated.
		}
		\label{fig:High}
	\end{figure}
\end{center}

\subsection{Two-photon spectroscopy at strong excitation}

In addition to the structure around the defect-qubit anticrossing, at higher power ($\Omega^{(1)}_q/h \approx7$~MHz) we find additional features at lower flux bias,
shown in Fig.~\ref{fig:High}. We observe a weaker line running parallel to the single-photon qubit transition,
showing an additional anticrossing with the $\ket{0g} \leftrightarrow \ket{1e}$ line,
indicated by the lefthand dashed box on the figure. From their power dependence, both these lines can be identified as two-photon transitions.

To explain the spectral features of Fig.~\ref{fig:High}(a), it is now essential to include the third level of the qubit circuit in the treatment.
A sketch of the level structure around this second anticrossing is again shown in the inset of Fig.~\ref{fig:High}.
Possible transitions are indicated by arrows. Due to the anharmonicity in the qubit circuit,
the transition between the levels $\ket{1} \leftrightarrow \ket{2}$ is detuned from the qubit transition $\ket{0}\leftrightarrow\ket{1}$ by the amount $\Delta = \epsilon_{12} - \epsilon_{01} < 0$.
At the chosen range of flux bias, our circuit has an anharmonicity $|\Delta|\sim 100$ MHz. We therefore identify this weaker,
parallel line as the two-photon $\ket{0g} \leftrightarrow \ket{2g}$ transition, detuned from the first transition line by the amount $|\Delta|/2 \sim 50$ MHz due to its two-photon nature.
The anticrossing in the left dashed box is due to the coupling between qubit and TLS $\sim \hat{p} \cdot \hat{\tau}_{x}$,
which hybridizes the states $\ket{1e}$ and $\ket{2g}$. The coupling strength is increased by a factor $\sqrt{2}$ due to the properties of the momentum operator
and the magnitude of the anticrossing is given by $ \sqrt{2} v_{\perp} / 2 \approx 15$ MHz (where the additional factor of two results from the two-photon process).
Fig.~\ref{fig:High}(b) shows the result of a numerical calculation. The theory is in very good agreement with the experimental data over all parameter regimes.

\begin{figure}[htb]
    	\includegraphics[width=0.35\columnwidth]{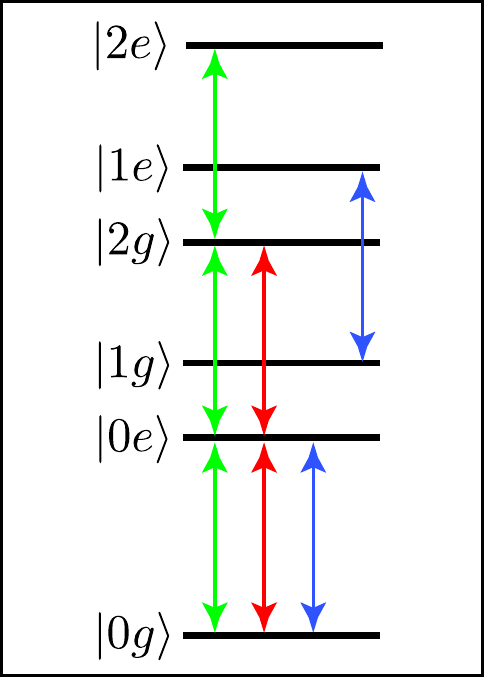}
    	\caption{(Color online) Level structure of the coupled system of qubit and TLS for $\omega_{d} \approx \epsilon_{f} \approx \frac{1}{2} \epsilon_{02}$, illustrating the high degree of symmetry. Possible microwave induced transitions at this driving frequency are indicated by arrows.
	}
	\label{fig:Blob}
\end{figure}

An additional feature can be identified in the spectroscopic scan of Fig.~\ref{fig:High}(a) at a flux bias of $1.175$
and microwave frequency resonant with the TLF $\omega_{d} \approx \epsilon_{f} \approx \frac{1}{2} \epsilon_{02}$. Here, the two-photon $\ket{0g} \leftrightarrow \ket{2g}$ line
crosses the one photon $\ket{0g} \leftrightarrow \ket{0e}$ transition. This is an area of high symmetry in the spectrum as is illustrated in Fig.~\ref{fig:Blob}, resulting in strongly enhanced absorption. The asymmetric nature of this effect is well reproduced by the numerics (Fig.~\ref{fig:High}(b)).

\subsection{Single photon spectroscopy of upper hybridised states}

\begin{figure}[htb]
    	\includegraphics[width=0.9\columnwidth]{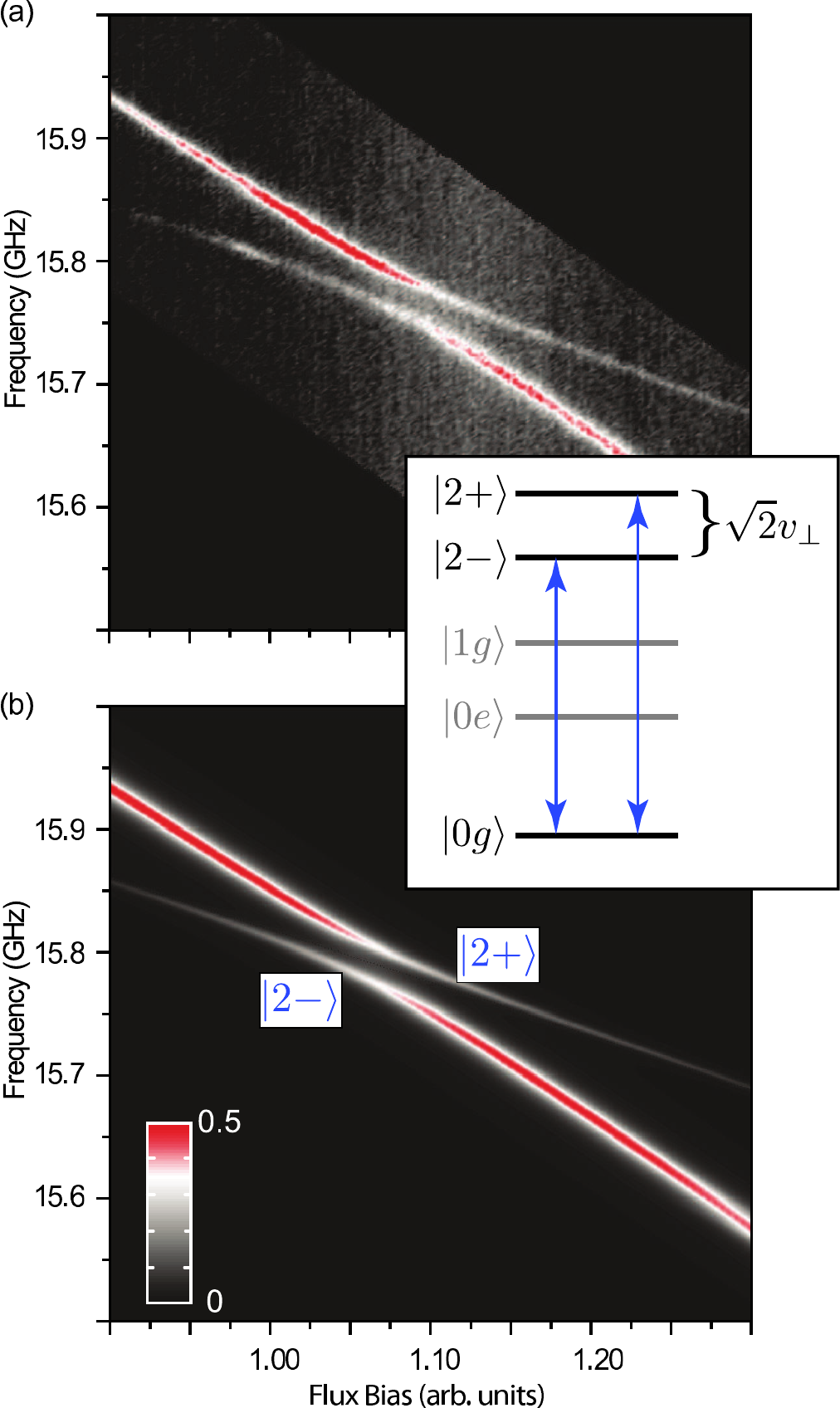}
    	\caption{(Color online) Using higher frequency excitations, the hybridised states $\ket{2\pm}$ can be directly excited via a one-photon process.
		Experimental \textbf{(a)} and theoretical \textbf{(b)} spectrum showing the clear anti-crossing due to the $\ket{0g} \leftrightarrow \ket{2\pm}$ transition. The colorscale is the same for both graphs.
		\textbf{(Inset)} The corresponding energy level diagram shows this direct transition which is typically ignored within the usual simplified models of such a system.
	}
	\label{fig:Two}
\end{figure}

As a further confirmation that the left anticrossing of Fig.~4 indeed represents the coupling between the states $\ket{1e}$ and $\ket{2g}$,
we probed these higher lying states directly with a single-photon transition. Such a process is equivalent to the cooperative absorption of single photons by pairs of rare-earth ions in crystals~\cite{Varsanyi:61}. The experimental results are presented in Fig.~\ref{fig:Two}(a) where one can clearly see the anticrossing at doubled resonance frequency $\sim 15.7$ GHz with a magnitude now of $\sqrt{2} v_{\perp} \sim 30$ MHz. When approximating the shallow well using the third order anharmonic oscillator potential~\cite{KofmanPRB06} the transition $\ket{0} \leftrightarrow \ket{2}$ is forbidden. However, when treating the anharmonicity exactly, there is a finite matrix element in the momentum operator $\hat{p}$ corresponding to this transition. Fig.~\ref{fig:Two}(b) shows the results of a theoretical calculation and the inset gives a sketch of the level structure including the transitions induced by the microwave drive.

\section*{V. Conclusion}

In conclusion, we have presented detailed data on experiments which probe the hybridization between a superconducting phase qubit
and a two-level defect in the tunnel barrier of its Josephson junction.
These elements are all strongly interacting, forming a regime analogous to strong coupling in circuit QED, with the role of the photon mode played by the TLS. The resulting hybrid quantum system is in turn probed with multi-photon spectroscopy. Understanding the experimental results requires a theoretical model which must go beyond the standard weak driving, harmonic well and two-level approximations.
Using this tight link between precision spectroscopy and systematic theoretical models,
we are able to map out the quantum mechanical nature of this hybrid system, combining elements from two length scales differing by four orders of magnitude.

\section*{Acknowledgments}

We would like to thank M. Ansmann and J. M. Martinis (UCSB) for providing us with the sample that we measured in this work. P. Bushev thanks P. Bertet and I. Protopopov for clarifying discussions. This work was supported by the CFN of DFG, the EU projects SOLID, EuroSQIP and MIDAS, and the US ARO Contract W911NF-09-1-0336.

\bibliographystyle{apsrev}
\bibliography{QuantumHybrid}

\end{document}